\documentclass{svjour3}                  

\usepackage{graphicx}
\usepackage{amssymb}
\usepackage{amsmath}
\usepackage[font=small, center]{caption}
\usepackage[separate-uncertainty=true,binary-units=true,per-mode=symbol]{siunitx}

\newcommand\cod[1]            {\texttt{#1}}    
\newcommand{\be}              {\begin{equation}}
\newcommand{\ee}              {\end{equation}}
\newcommand{\astri}           {ASTRI}
\newcommand{\astrih}          {ASTRI-Horn}
\newcommand{\miniarray}       {MiniArray}
\newcommand{\variance}        {Variance}

\newcommand{\fov}             {FoV}
\newcommand{\psf}             {PSF}

\DeclareSIUnit\mag{mag}

\journalname{Experimental Astronomy}

\begin{document}

\title{Assessment of the Cherenkov camera alignment through Variance images for the ASTRI telescope}

\titlerunning{ASTRI telescope: assessment of camera alignment through Variance images} 

\author{Simone Iovenitti   \and
        Giorgia Sironi     \and
        Enrico Giro        \and
        Alberto Segreto    \and
        Osvaldo Catalano   \and
        Milvia Capalbi}

\institute{Simone Iovenitti \at
              Università degli Studi di Milano \\
              INAF--Osservatorio Astronomico di Brera \\
              \email{simone.iovenitti@inaf.it} }

\date{Received: date / Accepted: date}

\maketitle

\begin{abstract}

A peculiar aspect of Cherenkov telescopes is that they are designed to detect atmospheric light flashes on the time scale of nanoseconds, being almost blind to stellar sources. As a consequence, the pointing calibration of these instruments cannot be done in general exploiting the standard astrometry of the focal plane.
In this paper we validate a procedure to overcome this problem for the case of the innovative ASTRI telescope, developed by INAF, exploiting sky images produced as an ancillary output by its novel Cherenkov camera. In fact, this instrument implements a statistical technique called ``Variance method'' (VAR) owning the potentiality to image the star field (angular resolution $\sim\SI{11}{\arcminute}$). We demonstrate here that VAR images can be exploited to assess the alignment of the Cherenkov camera with the optical axis of the telescope down to $\sim\SI{1}{\arcsecond}$. To this end, we evaluate the position of the stars with sub-pixel precision thanks to a deep investigation of the convolution between the point spread function and the pixel distribution of the camera, resulting in a transformation matrix that we validated with si\-mu\-lat\-ions. After that, we considered the rotation of the field of view during long observing runs, obtaining light arcs that we exploited to investigate the alignment of the Cherenkov camera with high precision, in a procedure that we have already tested on real data.\\
The strategy we have adopted, inherited from optical astronomy, has never been performed on Variance images from a Cherenkov telescope until now, and it can be crucial to optimize the scientific accuracy of the incoming \miniarray{} of ASTRI telescopes.


\end{abstract}

\section{Introduction}                      \label{sec:intro}

\astrih{} is a Cherenkov telescope entirely developed by the Istituto Nazionale di Astrofisica (INAF), aiming at the study of very-high energy (VHE) cosmic radiation, up to 100 TeV and beyond \cite{scuderi_astri_param}. It is a prototype instrument both for the realization of the \astri{} \miniarray{} observatory and for 
the opto-mechanical structure of Small-Sized Telescopes (SST) to be implemented in the Cherenkov Telescope Array (CTA) \cite{ASTRI_validation_2019}. \astrih{} is an example of Imaging Atmospheric Cherenkov Telescope (IACT) \cite{evolution_hillas}, designed to detect nanosecond light flashes due to the interaction of the incoming cosmic radiation with the atmosphere. In par\-ti\-cu\-lar, \astrih{} presents some peculiar novelties that make it a unique telescope in its genre, as it is briefly discussed hereafter. 

\noindent
\astrih{} is the first Cherenkov telescope with a dual-mirror configuration \cite{ASTRI_structure} (a modified Schwarzschild-Couder design \cite{astri_optical_validation}), allowing to have a wide, aplanatic field of view (\fov{}) up to \ang{10.5;;} diameter, with a primary mirror raw aperture of about \SI{4.3}{\meter}, despite a short equivalent focal lenght of $\sim$\SI{2.15}{\meter} \cite{ASTRI_DEFLECTOMETRY}. This optical scheme presents a small plate-scale (\SI{37.5}{\milli\meter\per\deg}), allowing to use miniaturized Silicon Photo-Multipliers sensors (SiPM) in the focal plane, sized on purpose to couple with the point spread function (\psf{}, $\sim\SI{7}{\milli\meter}$) of the instrument \cite{astri_optical_validation}. Each SiPM constitutes a square pixel covering $\sim$\ang{;11.2;} in the sky, with sensitivity in the range of wavelengths between \num{300} and \SI{550}{\nano\meter} with an appropriate filter. Globally, the whole camera is a mosaic of 37 tiles, called Photon Detection Modules (PDM), each composed of 64 pixels \cite{ASTRI_camera_2018}. 
However,
\astrih{} operated with a reduced PDM configuration of only 21 tiles, corresponding to $\sim$\ang{7.5;;}, in order to carry out the instrument performance validation.

\noindent
The primary output of the camera are the images of nanosecond atmospheric Cherenkov flashes, produced by the incoming VHE cosmic radiation. From this data stream, all the scientific information about astrophysical quantities can be retrieved, using the Hillas parametrization \cite{hillas_parameters} and suitable Monte Carlo simulations to analyze the shape of the flashes.
The stellar component of the night sky background light is almost invisible in this output, as the electronics was specifically designed to filter any steady or slow-varying incoming signal, as they are not of interest for the Cherenkov data reduction.
However, differently from other Cherenkov detectors, the \astri{} camera actually owns the potentiality to image the star field, with an ancillary output which is proportional to the incoming photon flux over every pixel, exploiting a statistical technique implemented in the logic board: the \variance{} method\footnote{
    The innovative Cherenkov camera CHEC \cite{CHEC}, recently developed, can also produce sky images exploiting a technique similar to the \variance{}, but using a separate DC coupled line to extract (integrate) the signal \cite{CHEC_POINTING}.
}.
It is based on the analysis of the electric signal detected by the camera front-end electronics, sampling randomly the output of the pixels not triggered by the Cherenkov events.
The net result is, for each pixel, a sequence of values whose average is constant with time, but whose variance is proportional to the flux 
on the pixel \cite{ASTRI_camera_2018}.
This technique is inherited from the photo-multipliers tube technology and with the \astri{} project it has been applied for the first time to a Cherenkov telescope equipped with SiPM sensors.
Using the \variance{} data, we can monitor the night sky background light and access the position of the stars in the field of view, opening the intriguing possibility to assess the actual pointing of the telescope without requiring any auxiliary optical instrument as every other IACT does \cite{Segreto_calibration}.

\begin{figure}[ht] 
\centering
  \includegraphics[width=\textwidth]{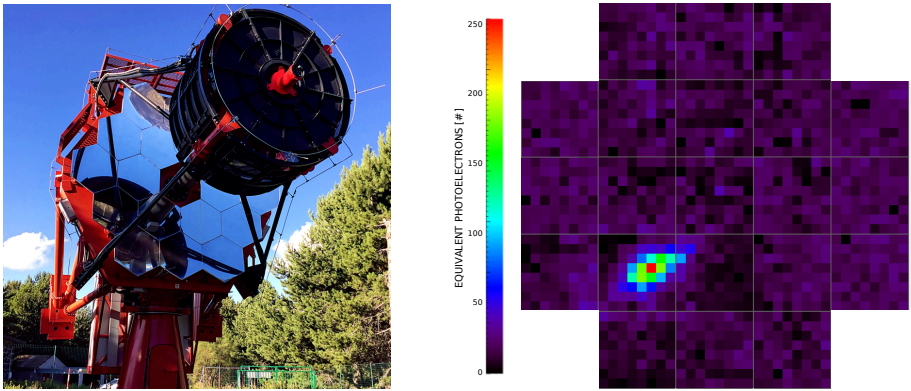}
\caption{The \astrih{} prototype telescope (left) and a typical Cherenkov event detected by the SiPM camera (right): no stars are visible in scientific images.}
\label{FIG_000}
\end{figure}

\subsection{Scope of this work}


The pointing direction of the \astri{} Cherenkov SiPM camera cannot be assessed through the use of images taken directly with it, because the acquisition electronics filters out the steady or slow-varying component of the sky light and hence the star field cannot be adopted as a reference.
For this reason, it is difficult to measure any eventual undesired effect introducing a mis-pointing of the camera system, e.g. mechanical tolerances or gravity flexures.
Despite this, in this paper we validate a method to measure the position of the camera geometric center with respect to the actual telescope optical axis pointing direction, exploiting the apparent motion of stars into \variance{} images.
To this end, the main challenge is to achieve the angular resolution necessary for scientific objectives (tens of arcseconds), despite the large pixel size (few arcminutes) and the non-uniform gaps between the tiles of the Cherenkov camera \cite{ASTRI_camera_2018}. To overcome these problems, we adopted a very common approach: to study the behaviour of the \psf{} as it crosses the pixels during the apparent rotation of the field of view due to the sideral motion and the alt-azimuthal mount of the telescope \cite{Segreto_calibration}. This technique is often adopted in optical astronomy \cite{EXPLOIT_FOV_ROTATION}, but it has never been performed on data from a Cherenkov camera, until now.

\begin{figure}[ht] 
\centering
  \includegraphics[width=0.8\textwidth]{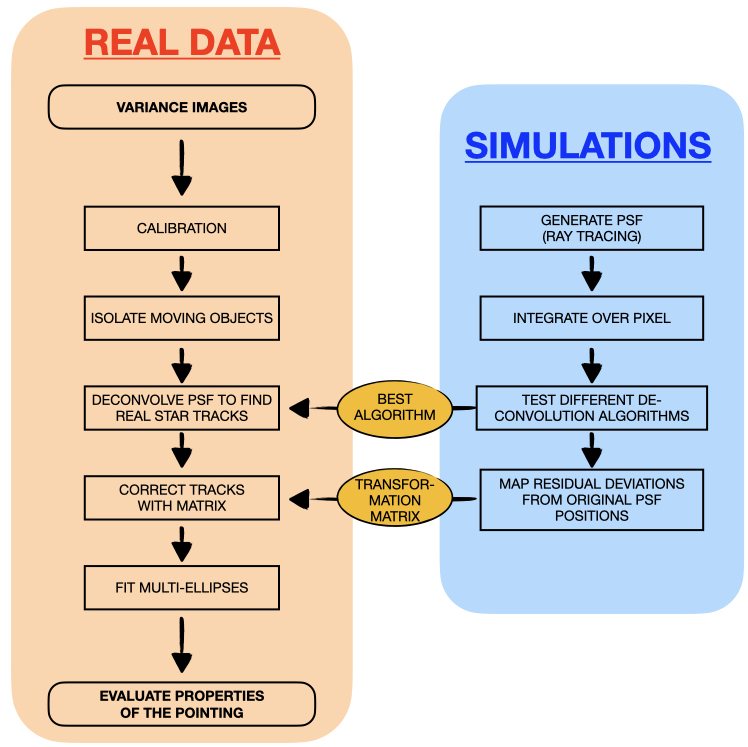}
  \caption{Flowchart of the procedure presented in this paper.}
\label{Fig_flowchart}
\end{figure}

\noindent
In this work we demonstrate that, in the case of \astri{}, provided at least 4 bright stars ($ \lesssim \SI{5.5}{\mag} $) in the \fov{} it is possible to verify the alignment of the Cherenkov camera with the optical axis of the telescope with a precision down to $\sim\SI{1}{\arcsecond}$.
To this end, we developed a pipeline to analyse the star trails (section~\ref{sec:analysis}) using an algorithm to retrieve the position of stars with sub-pixel accuracy (section~\ref{sec:convolution}). We validated our procedure through simulations, then we applied it to real data taken with \astrih{}.
Figure~\ref{Fig_flowchart} shows a flowchart of the whole procedure, while section~\ref{sec:conclusion} contains a discussion of the results, together with the future improvements to implement it in the \astri{} \miniarray{}.\\
To begin, section~\ref{sec:clusters} starts with the sample selection.

\section{Data selection and pre-processing}          \label{sec:clusters}

\variance{} data are taken in parallel to scientific data, during every observing run of the telescope. In particular, the electronics of \astrih{} was configured to generate a \variance{} image, or \textit{frame}, every $\sim$\SI{3.25}{\second}.
In these images, the brightest stars ($ \lesssim \SI{5.5}{\mag} $) are clearly visible and they present the typical \fov{} rotation of every telescope with an alt-azimuthal mount performing a tracking observation, as it is reported in the next section.
In order to analyse the trajectories of the stars in the \fov{}, only the observing runs with the largest rotation must be selected among the data collected by \astrih{}\footnote{
   As a prototype telescope, \astrih{} accomplished only one full observing campaign, pointing at the Crab Nebula region, taking data in ON/OFF mode \cite{ASTRI_CRAB_detection}.
}. After that, images must be calibrated with a flat-field-like procedure. Finally, pixels containing stars must be isolated for the detailed analysis.

\subsection{Rotation of the field of view}       \label{sec:rotation}
The angular extension of the \fov{} rotation can be calculated considering that it is given by the evolution of the \textit{parallactic angle} $\eta$, i.e. the angle between the north celestial pole and the zenith, with the vertex in the pointing direction \cite{SphericalAstronomy}.
At any time, during a tracking observation of an object, the value of $\eta$ is given by \cite{PAR_ANGLE}
\be  \frac{\sin \eta}{\sin \left( \frac{\pi}{2}-\phi \right) } = \frac{\sin h}{\sin z}\;,      \label{eq:parangle}                     \ee
where $\phi$ is the telescope’s latitude, $h$ is the object’s hour angle and $z$ is the object’s zenith angle, that we can retrieve from the position of the telescope motors' encoders. The time dependence is embedded into $h$ and $z$, and there is not a simple relation between the amplitude of the rotation $\Delta\eta$ and the time length of the observing run\footnote{
For simulating in details the \fov{} rotation we adopted the software ``Star Coverage'' \cite{IOVE_ICRC21_STARCOVERAGE}.
}.
To test our procedure, we selected those cases where the telescope was in tracking mode and $\Delta\eta \geq \ang{25;;}$ among all the data sets available in the \astrih{} archive: we identified 9 observing runs\footnote{
In principle, consecutive runs on the same target can be concatenated for our analysis, thus increasing the total angular coverage, but for testing the procedure we decided to work on \textit{single} runs only.
}. Among them, only 2 cases present at least 4 bright stars in the \fov{} and we chose them to test our procedure. Table~\ref{Tab:runid} report their parameters.  
\begin{table}[h!]
\centering
	\begin{tabular}{ c c c c c c}
	\hline
	\noalign{\smallskip}
	{Run ID}	&	{Date} 	&   {Pointing}  &  {Frames}  &  {Time lenght}   &  {Angular coverage}  	\\
	\noalign{\smallskip}
	\hline
	1597	& 2019-02-26	 &	Crab Nebula  &  2275  &   1h 59m 44s  &   \ang{33.4;;}       \\
	1620	& 2019-02-28	 &	Crab Nebula  &  1986  &   1h 43m 39s  &   \ang{25.5;;}        \\
	\hline
	\end{tabular}
\caption{Parameters of observing runs that we considered as test-case. }
\label{Tab:runid}
\end{table}

\begin{figure}[ht] 
\centering
  \includegraphics[width=\textwidth]{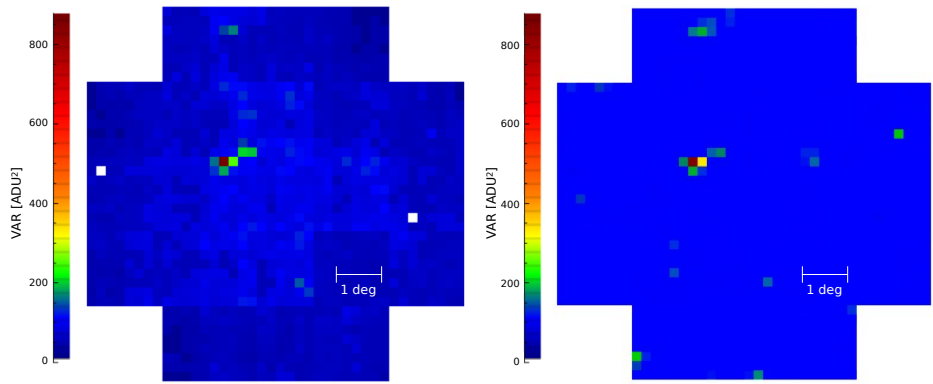}
\caption{Visualization of the \variance{} data taken with the \astrih{} Cherenkov camera before (left) and after (right) the flat-field calibration. Local maxima are stars in the \fov{}, the brightest one is \textit{Tianguan} or Zeta Tauri (123 Tau). Observing run 1597 (see table~\ref{Tab:runid}).}
\label{Fig_0_calibration}
\end{figure}

\subsection{Gain calibration}       \label{sec:calibration}

The pixel values in the \variance{} images must undergo a gain-calibration process (flat field) whose parameters has unfortunately not been characterized in laboratory yet\footnote{
    By now, the full calibration of the camera was performed only for scientific data \cite{CAMERA_CALIBRATION}.
}. Thus, we performed the calibration adopting a statistical approach: for every pixel, we consider the median \variance{} level in every run (in order to mitigate the effect of the star spots occasionally transiting over some pixels in some frames) and then we take the average of these values over our sample of 9 long observing runs. Hence, the calibration coefficient $C_i$ for every pixel $x_i$ is given by
\be                   C_i = \frac{1}{N} \sum_{\mathrm{RUNS}} \mathrm{median}(x_i, f)  \;,    \label{eq:calibration}              \ee
where $f$ is the set of frames composing every observing run and $N$ is their number. The final calibration factor $F_i$ is given by the inverse of $C_i$, after an appropriate normalization, and the pixel value in every frame is finally calculated with $x_i\times F_i$.\\
Figure~\ref{Fig_0_calibration} shows the effects of this calibration.

\subsection{Identify the stars}      \label{sec:tracks}   

\begin{figure}[ht] 
\centering
  \includegraphics[width=0.6\textwidth]{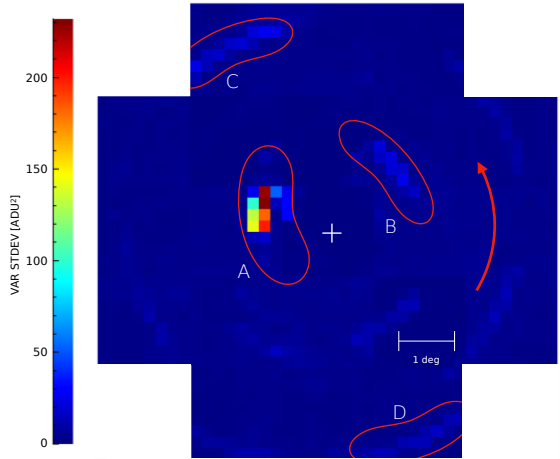}
  \caption{Star trails isolated by the clustering algorithm. Camera center is pointing the Crab Nebula.
           Star A is 123 Tau (\textit{Tianguan}, \SI{3.0}{\mag}),
           B is 114 Tau (\SI{4.8}{\mag}),
           C is 119 Tau (\SI{4.3}{\mag}) and
           D is 125 Tau (\SI{5.2}{\mag})
           [values taken from SIMBAD, V band].}
\label{Fig_1_cluster}
\end{figure}

During the rotation of the \fov{} the image of every star changes position on the camera, thus increasing the ``variance of the \variance{}'' of the pixels receiving light. As a consequence, considering for each pixel the standard deviation of the whole observing run, we can visualize the star trails, as it is shown in figure~\ref{Fig_1_cluster} for the observing run 1597. After that, a clustering algorithm (centroid-based, of the \emph{k-means} family \cite{cluster_analysis}) isolates the groups of pixels belonging to the same moving spot, hence identifying the tracks of the brightest stars.
In the following part of the analysis, only pixels within the clusters will be considered to retrieve the position of the stars in every frame.

\section{Retrieve the position of stars}      \label{sec:convolution}

\begin{figure} 
\centering
\includegraphics[width=0.6\textwidth]{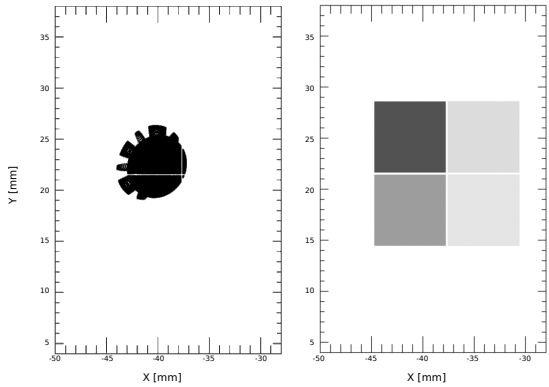}
    \caption{ Ray-tracing simulation of 83518 photons from a point-like source \ang{1.25;;} off the optical axis of the telescope. \textit{Left:} the resulting \psf{} of the \astrih{} telescope. \textit{Right:} the convolution of that \psf{} over the Cherenkov camera pixels. The gray scale is proportional to the number pf photons collected by every pixel, while the white separation lines represent the gaps between them.}
\label{Fig_2_psf_pixel}
\end{figure}

In order to draw the track of the stars in the \fov{} with sufficient precision (few arcminutes),their position in every frame must be determined with sub-pixel accuracy. Unfortunately, the shape of the \psf{} cannot be exploited to this aim, because its average extension is almost equal to the pixel size ($\sim$ \SI{7}{\milli\meter}) \cite{ASTRI_validation_2019}. However, when the star spot falls across multiple pixels, it is possible to retrieve the position of the centroid considering their different illumination, i.e. the \psf{} convolution over the pixel distribution (see figure~\ref{Fig_2_psf_pixel}). To this end, we simulated the pixel illumination by a generic celestial point-source and we tested different de-convolution algorithms using the dispersion from the original star position as a figure of merit (section~\ref{sec:barycenter}). Moreover, we improve this result studying in detail the effect of the gaps between the pixels (section~\ref{sec:distortion}).

\subsection{Deconvolution of the PSF}       \label{sec:barycenter}
In order to choose the best algorithms to de-convolve the \psf{} from the \variance{} images, we started simulating the whole process of image creation in the \astri{} camera. We artificially generated a celestial point-source off axis and performed a full ray-tracing of \num{83518} photons to produce its theoretical spot onto the camera\footnote{
    The procedure to simulate the \psf{} was validated with direct measurements during the commissioning phase of the \astrih{} telescope \cite{ASTRI_validation_2019}. }.
Then we considered its apparent motion due to the evolution of the parallactic angle $\eta$, given by equation~\ref{eq:parangle}. We integrated light spots over the camera pixel area, obtaining a \variance{}-like set of images. Finally, we tested different strategies to recover the position of the original spot centroid, evaluating their accuracy via
the standard deviation $\delta$ of the reconstructed points with respect to the initial simulated trail\footnote{$\delta$ is measured in \si{\milli\meter} and converted in arc-minutes using the plate-scale.}.
\begin{itemize}

    \item
    \textit{Gaussian fit (Gauss),} $\delta = \ang{;2.39;}$. 
    The \variance{} value of every pixel is assigned to the position of the pixel center and the resulting matrix is interpolated to obtain an image; every local maximum is fitted with a 2D-Gaussian function to obtain the centroid position.
    
    \item
    \textit{Weighted-average  (W.A.),} $\delta = \ang{;0.81;}$. 
    Within each cluster (see sec.~\ref{sec:tracks}), we consider the pixel with the maximum value together with its eight nearest neighbor\footnote{
    Using 9 pixels in total we are sure to consider also the tails of the \psf{}.}: the barycenter of the simulated light spot is given by the average of pixels center positions, weighted by the recorded intensities.
    
\end{itemize}

\begin{figure} 
\centering
  \includegraphics[width=0.6\textwidth]{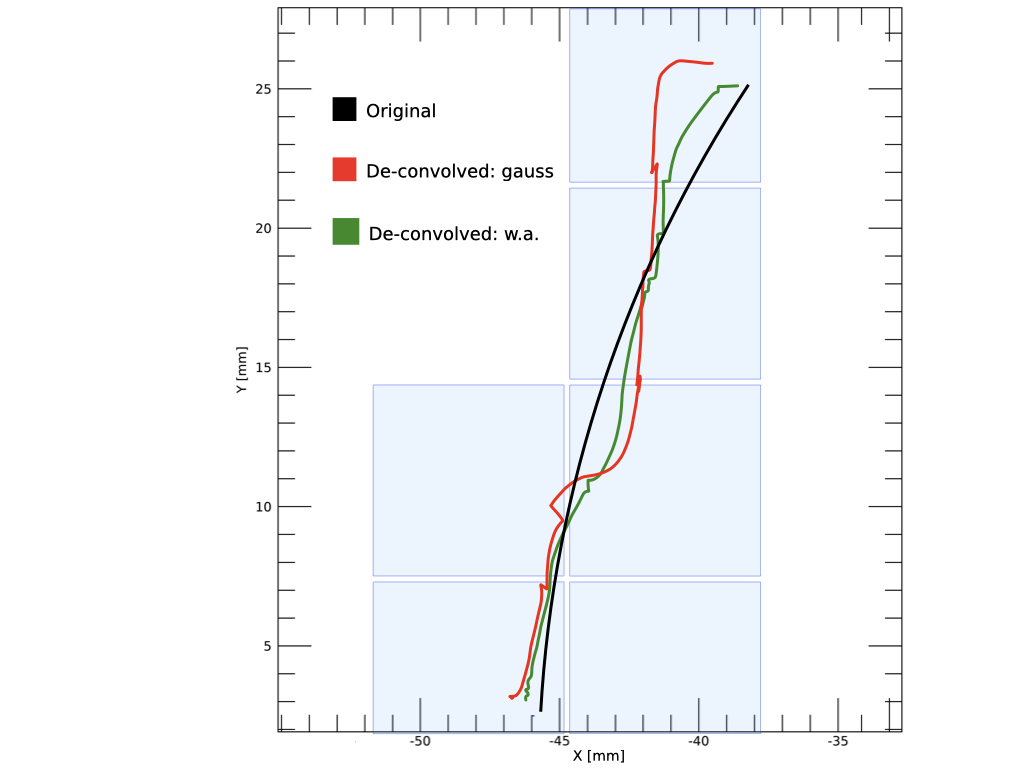}
  \caption{ Track of the \psf{} centroid of the original simulated star (black) and its reconstruction with different algorithms after the integration of light over the pixels. The time length of the simulation is \SI{2}{\hour}. Light blue squares represent six pixels of the camera.}
\label{Fig_3_convolution}
\end{figure}

\noindent
Figure~\ref{Fig_3_convolution} shows the outcome of the two methods described: W.A. presents a better match with the original curve, so this method was adopted, as it requires fewer further corrections to retrieve the original star trail. In addition, figure~\ref{Fig_3_convolution} also presents little jumps close to the gaps between the pixels, suggesting that their role must be studied in details: they will be considered in section~\ref{sec:distortion}.  

\subsection{Effects of gaps between pixels}            \label{sec:distortion}
\begin{figure} 
\centering
    \includegraphics[width=1.0\textwidth]{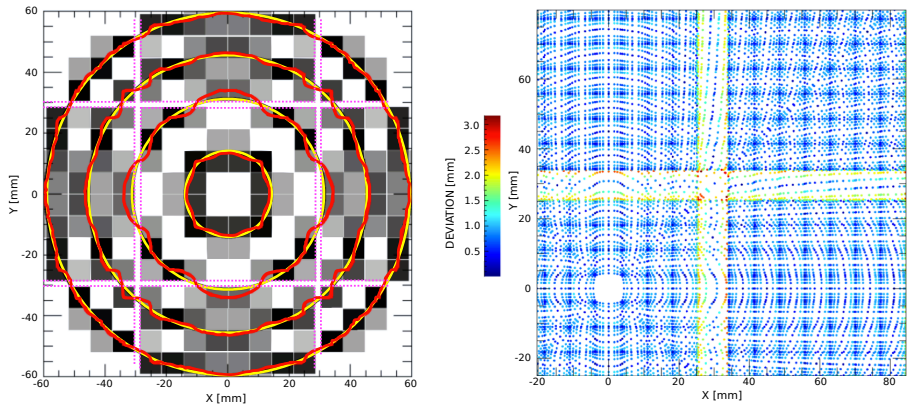}
    \caption{ Left: central portion of the \astrih{} camera with the simulation of the tracks of 4 stars rotating along perfect circles (\ang{360;;}).
              The reconstruction of the tracks (red line) presents considerable deviations from the theoretical curve (yellow) when the spot crosses the gap between PDMs (magenta).The grey scale of the pixels is proportional to their illumination.
              Right: each point represents the position of a star centroid retrieved by our algorithm and the color scale expresses its displacement from the theoretical position.}
\label{Fig_4_5}
\end{figure}

Gaps between pixels subtract photons to the star spots, modifying the shape of the \psf{} and distorting the outcome of the deconvolution algorithm\footnote{
The loss of photons is negligible for the analysis of the Cherenkov showers, as they typically cover a total surface made of several pixels, while dead areas are very small in comparison.}.
This effect is enhanced at the edge of PDMs, where the gap is larger (\num{0.8} to \SI{1.6}{\milli\meter}) compared with the gap between ``internal'' pixels (\SI{0.2}{\milli\meter}) \cite{TN008}. Figure~\ref{Fig_4_5} highlights this phenomenon, with a virtual \fov{} rotation of \ang{360;;}.

\noindent
In order to characterize and remove the effect of pixel gaps, we simulated a \psf{} in every point of a fine grid of possible positions in the camera and we measured the distance between the original spot centers and the ones calculated with the deconvolution algorithm \textit{(W.A.)}. We obtained a map of the distortion introduced by the geometry of the camera (reported in figure~\ref{Fig_4_5}, right) that can be used as a transformation matrix to correct the position of calculated centroids. The effectiveness of this procedure is verified applying the transformation matrix on the track of figure~\ref{Fig_4_5}, left (red line).
The maximum deviation $\delta$ is reduced of about $\SI{40}{\percent}$.

\section{Analysis of the real tracks observed by \astrih{}}                       \label{sec:analysis}

To take into account any possible deviation from the expected pattern, 
we assumed that the shape of the real star trails can lie along multiple arcs of ellipses with the same center, eccentricity and tilt angle, as it is shown in figure~\ref{Fig_6_ellissi}. To extract the parameters of the ellipses we fitted the data with a suitable function (section~\ref{sec:fit}). Furthermore, we investigated the dependency of the results on the angular coverage, determining in which conditions our procedure can ensure the highest precision (section~\ref{sec:discussion}).

\subsection{The fit procedure}            \label{sec:fit}
\begin{figure} 
\centering
    \includegraphics[width=\textwidth]{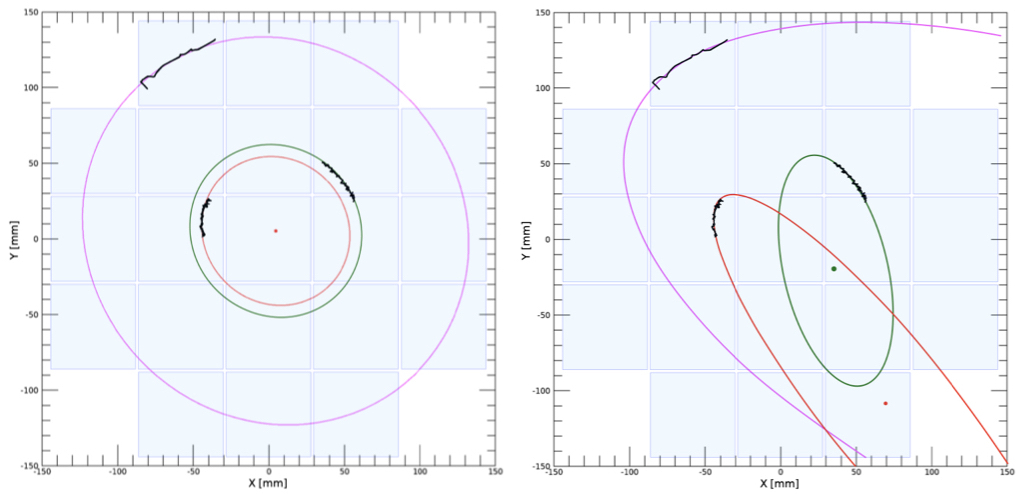}
    \caption{ Output of the fit described in section~\ref{sec:fit}: star tracks from real \variance{} data are represented in black while light-blue squares represent PDMs of the \astrih{} camera. \textit{Left}: 3 stars were considered simultaneously for the fit; \textit{right}: each star was fitted individually.}
\label{Fig_6_ellissi}
\end{figure}

The trails of the stars' centroids are fitted with a function that we defined on purpose,
describing simultaneously multiple ellipses, with the same center, eccentricity and tilt angle. The fit was performed with the \cod{MPFIT} procedure of the \cod{IDL} library \cite{IDL_MPFIT}. The number of free parameters is $p=4+n$, where $n$ is the number of stars. In particular, the parameters of the fitting functions are
\begin{itemize}
    \item $X_0$ and $Y_0$ of the center;
    \item $\theta$ tilt angle;
    \item major and minor semiaxes of the innermost elliptical star trail;
    \item major semiaxis of the other elliptical star trails.
\end{itemize}
The fit function is called several times, with random changes to the initial parameters and without constraints: the dispersion of the results is 
adopted as an estimation of their uncertainty.
Figure~\ref{Fig_6_ellissi} reports the mean of fit results for the observing run 1597, with 3 stars considered simultaneously in the trail fitting procedure. In this example, the center of the ellipse is in $X_0 = \SI{4.51(6)}{\milli\meter}$, $Y_0 = \SI{5.25(37)}{\milli\meter}$, equivalent to an angular distance of $\delta_0 = \SI{10.6(5)}{\arcminute}$ from the geometric center of the camera\footnote{
   Uncertainties in the $x$ and $y$ directions are evaluated independently: the error on $X_0$ is smaller as the data-set considered here provides a better coverage of the horizontal direction.
}. However, considering four stars for the a\-nalysis,
hence constraining the fit, 
the results becomes consistent with the geometric center of the camera, with an error lower than \ang{;;1}.
This behaviour is shown and discussed in the next section.

\subsection{Results}            \label{sec:discussion}
\begin{figure}[ht] 
\centering
  \includegraphics[width=\textwidth]{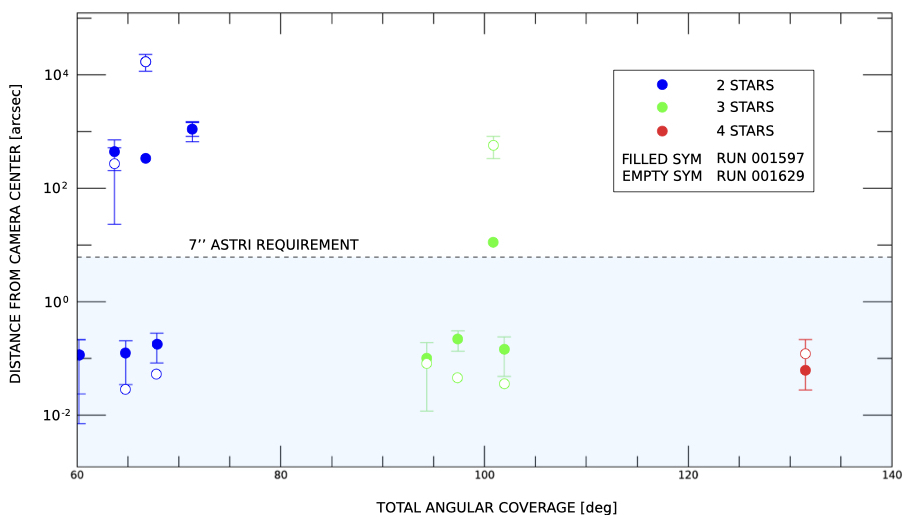}
\caption{Dispersion of fits results and their error, for combinations of 2,3 and 4 stars, from the two observing runs of table~\ref{Tab:runid} (denoted with filled and empty symbols). In abscissa there is the total angle covered by the sum of the stars considered for the fit during \fov{} rotation. The \SI{7}{\arcsecond} value for the pointing precision requirement of a single \astri{} telescope \cite{TN005} is reported as a reference. }
\label{Fig_7_dispersion}
\end{figure}

The dispersion of fit results strongly depends on the number of stars considered and their angular coverage, as it is reported in figure~\ref{Fig_7_dispersion}.
In particular, when less than 4 stars are included, the fit result is influenced by the edgy shape of the star-trail reconstruction, due to the large pixel size.
An example is reported in figure~\ref{Fig_6_ellissi}~\textit{right}, where the stars are fitted individually and the outcome is non-realistic, too far from the expected circular shape.
We investigated this effect performing our analysis on the observing runs selected in table~\ref{Tab:runid}, both presenting 4 bright stars, allowing us to consider different combinations of 2 stars (6), 3 stars (4) and 4 stars (1), per each of them. Results are reported in figure~\ref{Fig_7_dispersion}, showing the distance of their calculated centroids from the camera geometric center, with their errors. Table~\ref{Tab:parametri} reports a summary of this analysis.

\begin{table}[h!]
\centering
	\begin{tabular}{ c c c}
	\hline
	\noalign{\smallskip}
	{Stars [num]}	&	{Distance from center [\si{\arcsecond}]} 	& {Error [\si{\arcsecond}]}	\\
	\noalign{\smallskip}
	\hline
	2		& 1689	 &	620	         \\
	3 		& 73  	 &	33           \\	
	4   	& 0.1    &	$<$ 1.0      \\
	\hline
	\end{tabular}
\caption{Averages of fits results and their error, for the analysis with 2, 3 and 4 stars, on the real data from table~\ref{Tab:runid}.}
\label{Tab:parametri}
\end{table}

\noindent
It emerges that it is sufficient to have 4 bright stars in the \fov{} (like in the case of the Crab Nebula region) to obtain a  measure of the offset between the telescope axis and the geometric center of the Cherenkov camera with an uncertainty lower than \ang{;;1}. 
In fact, with 4 stars the fit procedure is well constrained and hence reliable, providing us a result which is exploitable for scientific purpose.

\section{Conclusion}                        \label{sec:conclusion}

\variance{} images represent a unique opportunity to study the sky portion actually framed by the Cherenkov camera of the \astrih{} telescope, owning the potentiality to finely calibrate and monitor its pointing \cite{Segreto_calibration}. In particular, the procedure presented in this paper reveals any possible offset in the alignment of the Cherenkov camera with the optical axis of the telescope, exploiting the technique of the \fov{} rotation that was never applied before to real images from a Cherenkov camera. We validated our method with simulated star trails and, analysing real data taken by \astrih{}, we demonstrated that its sensitivity is high enough to be exploited for scientific purpose, reaching the scale of arc-seconds if the \fov{} rotation is greater than \ang{25;;} and 4 bright stars ($\lesssim$5.5 mag) are present.

\noindent
This work represents the first scientific analysis performed on real \variance{} data from an IACT and constitutes a powerful diagnostic tool to assess its pointing properties, hence optimizing the scientific accuracy of the whole system.

\noindent
Lastly, we want to point out that applying our technique on a wider data sample by \astrih{} one could in principle verify the alignment of the Cherenkov camera during all the different campaigns of the observing period\footnote{
  We remind that \astrih{} is a prototype telescope, and often extraordinary maintenance operations were performed between different observing campaigns, also regarding the camera mount system.
}, possibly enhancing the fine pointing reconstruction \textit{post-facto}.
On the other hand, in the incoming \miniarray{} of \astri{} telescopes, the procedure presented in this work will be implemented since the first calibration phase, assessing the alignment of the camera mount systems from the beginning \cite{ASTRI_ICRC21_MINEO}.

\subsection{Possible limitations of this method}
At a first sight, the procedure we have presented for the alignment of the Cherenkov camera seems to require some very lucky conditions: no clouds along all the path of stellar sources during long acquisitions and a pretty crowded star field in the \fov{} in order to have a reliable reconstruction of the rotation. Actually, in real use cases passing clouds are not a problem, as it is not necessary that long star trails are continuous: several consecutive segments can be attached together and the fit procedure will be well constrained regardless to possible "holes" in the middle. On the other hand, to have 4 bright stars at least in the \fov{} is not so common, despite the large aperture of the \astri{} telescope. However, it is essential to point out that this alignment procedure of the camera will not be performed every night, but only on dedicated calibration sessions or in case of necessity \cite{ASTRI_ICRC21_MINEO}, and suitable sky regions will be chosen in those occasions.

\noindent
Regardless to the pointing direction, nowadays the sky view is full of bright satellites. Their presence can alter the reconstruction of the stars' position in our procedure, but fortunately in a negligible way. In fact, even high altitude satellites are very fast compared to the diurnal motion, and hence their light can affect the reconstruction of each star centroid for no more than ten point maximum along its track.

\noindent
Possible important limitations of our method are represented by distortions in the star trails introduced by external agents, e.g. strong wind gusts (but the weather station would record their critical values), or the presence of errors in the tracking (for example a drift due to errors in the pointing model or to the motors inertia): this effect would influence the result of our analysis in a way that is difficult to simulate in advance. To overcome this problem the first solution is to execute the calibration with our procedure when the telescope in staring-mode pointing at Polaris, avoiding artifacts introduced by the movement of the structure. Secondly, another solution is to monitor constantly, frame by frame, the pointing direction of the telescope with very high precision, so to exclude possible tracking errors. To this end, we have already developed\footnote{
    Paper in preparation.
} a tool to analyze the position of known stars in the \fov{} and to compare the \variance{} data with the log of the motors' encoders and the images from an optical auxiliary device: the Pointing Monitoring Camera (PMC)\footnote{
    An auxiliary optical camera installed behind the secondary mirror, on the optical axis of the telescope (compatibly with gravity flexions and mechanical tolerances).
}. This will ensure a robust and continuous monitoring of the pointing performances, so that the alignment of the Cherenkov camera can be investigated at a deep level with the procedure presented here, both during the assembly verification phase and dedicated calibration sessions.


\begin{acknowledgements}
We remember that this work is supported by the Italian Ministry of Education, University, and Research (MIUR) with funds specifically assigned to the Italian National Institute for Astrophysics (INAF). This article has gone through internal ASTRI review and we want to thank all the members of that commission. In particular, we would like to thank Giovanni Pareschi, PI of the experiment, for the constant supervision and for his helpful corrections in the manuscript. 
\end{acknowledgements}

\bibliographystyle{spmpsci}  
\bibliography{biblio} 

\end{document}